\documentclass[showpacs,twocolumn,preprintnumbers,amsmath,amssymb,prb,aps]{revtex4}
\usepackage{txfonts}
\usepackage{pifont}
\usepackage{amssymb}
\usepackage{booktabs}
\usepackage{graphicx}%
\usepackage{dcolumn}
\usepackage{amsmath}
\usepackage{epstopdf}
\usepackage{bm}

\hfuzz=\maxdimen
\makeatletter
\def\btt#1{\texttt{\@backslashchar#1}}%
\DeclareRobustCommand\bblash{\btt{\@backslashchar}}%
\makeatother

\begin{document}
\title{From a normal insulator to a topological insulator in plumbene}
\author{Xiang-Long Yu$^{1,2}$}
\author{Li Huang$^{1}$}
\author{Jiansheng Wu$^{1}$}
\email[Corresponding author. E-mail: ]{wujs@sustc.edu.cn}
\affiliation{1 Department of Physics, South University of Science and Technology of China, Shenzhen 518055, P.R. China}
\affiliation{2 School of Physics and Technology, Wuhan University, Wuhan 430072, P.R. China}

\begin{abstract}
  Plumbene, similar to silicene, has a buckled honeycomb structure with a large band gap ($\sim 400$ meV). All previous studies have shown that it is a normal insulator. Here, we perform first-principles calculations and employ a sixteen-band tight-binding model with nearest-neighbor and next-nearest-neighbor hopping terms to investigate electronic structures and topological properties of the plumbene monolayer. We find that it can become a topological insulator with a large bulk gap ($\sim 200$ meV) through electron doping, and the nontrivial state is very robust with respect to external strain. Plumbene can be an ideal candidate for realizing the quantum spin Hall effect at room temperature. By investigating effects of external electric and magnetic fields on electronic structures and transport properties of plumbene, we present two rich phase diagrams with and without electron doping, and propose a theoretical design for a four-state spin-valley filter.
\end{abstract}


\maketitle

\section{INTRODUCTION}
In $2005$, Kane and Mele first proposed the prototypical concept of quantum spin Hall (QSH) insulator in graphene\cite{PRL95.146802,PRL95.226801}. Although the spin-orbit coupling (SOC) opens a band gap at the Dirac point, the associated gap is so small ($ \sim 10^{-3}$ meV\cite{PRB75.041401}) that the QSH effect in graphene only appears at an unrealistically low temperature. Subsequently, other group-IV materials with a honeycomb-like lattice and larger SOC interaction are reported to host the QSH effect with experimental accessible gaps, such as silicene, germanene and stanene\cite{PRL102.236804,PRL107.076802,PRB84.195430,PRL109.016801,PRL113.256401,PRB90.121408,RPP79.066501}. However, their gaps still are not large enough for the QSH effect which can be realized easily at room temperature. Up to now, the quantized conductance through the QSH edge states have only been verified in several experiments\cite{RPP79.066501} (\textit{e.g.} Bi bilayer film\cite{PRL109.016801}, HgTe/CdTe\cite{Science318.766}, and InAs/GaSb\cite{PRL114.096802} quantum wells). These QSH experiments have various limitations, such as a small bulk gap, toxicity and technological difficulties during processing. The interference of thermally activated carriers in the bulk is the main question for realizing the QSH effect at room temperature. Therefore, it is urgent to find new topological insulators with a large bulk energy gap that can stabilize the edge current.

At present two-dimensional ($2$D) materials in groups IV and V are promising candidates for the QSH effect. Theoretical calculations predict that a large bulk gap can be obtained through fluorination, hydrogenation, or other types of functionalizations\cite{PRL111.136804,PRB89.115429,NJP17.083036,SR5.8426,SR6.20152,Nano7.1250037}. The purpose of atom or molecule absorption is to saturate the $p_{z}$ orbital or dangling bond so that a band gap can be opened at the Dirac point without SOC. When the SOC is considered, a larger gap can be obtained in the corresponding system. For example, first-principles calculations predict a $2$D QSH insulator in F-decorated plumbene monolayer with an extraordinarily giant bulk gap of $1.34$ eV\cite{SR6.20152}. Unfortunately, a recent experiment has revealed that defects and disorders increase rapidly during plasma fluorination and hydrogenation of materials and even under short plasma exposures\cite{JACS133.19668}. The systems with chemical functionalizations may not be stable for long periods of time in air, especially with hydrogenation\cite{JACS133.19668,NJP17.083036}. Thus, it is rather difficult and challengable to prepare high-quality samples with chemical functionalizations, resulting in the unrealizable observation of the QSH effect in such experiments. For device applications it is almost impossible.

Therefore, to find QSH insulators with simple lattice structures and large bulk gaps has become our main purpose. Bi monolayer film is a promising candidate, and theoretical studies have also shown that it is a topological insulator with a large bulk gap ($\sim 500$ meV)\cite{PRL107.136805,PRL97.236805,PRB83.121310,SR5.11512,NJP16.105018}. But so far the QSH effect and topological properties of Bi monolayer have not been reported experimentally. Our present research focuses on another large-gap insulator material, plumbene monolayer. Recently the stability of the buckled $2$D Pb thin film was reported\cite{NJP16.105018,PRB90.241408,NC4.1500}. Previous theoretical works have given the same result that plumbene is a normal band insulator\cite{SR6.20152,NJP16.105018}. However, we find that the system can become a topological insulator through electron doping, which is feasible using gating methods
by electric fields in experiments\cite{APL90.052905,NN10.270}.
In this paper, we perform detailed first-principles and tight-binding calculations to investigate band structures and topological properties of plumbene.
Our results show that the band structures of gated plumbene harbour two spin-polarized Dirac cones with up to $96 \%$ ratio at the corner of the Brillouin zone, and that the system can change from a normal insulator with a large gap of $\sim 400$ meV to a topological insulator with a large bulk gap of $\sim 200$ meV. The QSH state is very robust with respect to external strain.
By investigating the effects of external perpendicular electric and magnetic fields on plumbene, two rich phase diagrams are obtained and we further propose a design for a four-state spin-valley filter. Owing to the simple structure, two large bulk gaps and various responses in external fields, plumbene can be a promising platform for topological phenomena and new quantum device applications at room temperature.

The remainder of the paper is organized as follows. In Sec. II, we describe the crystal structure of plumbene and our calculation methods, including first-principles methods and a sixteen-band tight-binding model. In Sec. III, we study the topological characteristics of plumbene without and with electron doping; through investigating the effects of external electric and magnetic fields on electronic structures of plumbene, we present two rich phase diagrams and a theoretical design for a four-state spin-valley filter is proposed; substrate effects on plumbene are also discussed. Finally, we conclude this work with a brief summary in Sec. IV.

\section{Computational details}
\subsection{Crystal structure}
\begin{figure}[tbp]
  \centering \includegraphics[width=8.0cm]{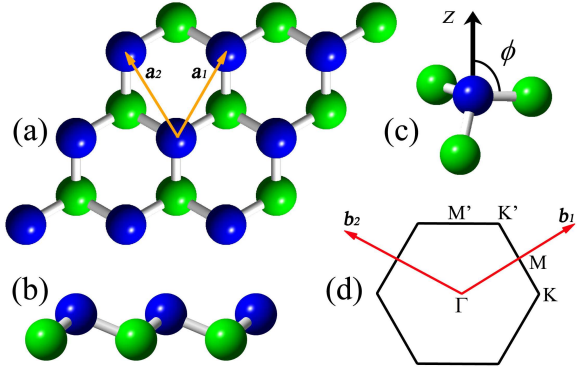}
  \caption{(Color online) Lattice structure of the buckled plumbene from (a) the top view and (b) the side view. (c) The angle between the Pb-Pb bond and the $z$ direction is defined as $\phi$. (d) The first Brillouin zone with hight-symmetry points. Green and blue spheres correspond to $A$ and $B$ sublattices, respectively. $\mathbf{a_{1}}$ and $\mathbf{a_{1}}$ are primitive vectors. $\mathbf{b_{1}}$ and $\mathbf{b_{1}}$ are reciprocal lattice vectors. }
  \label{fig:LattiStruct}
\end{figure}

Plumbene has a buckled honeycomb structure, as shown in Fig. \ref{fig:LattiStruct}. The lattice constant $a_{0} = 4.93$ \AA \ and the nearest-neighbor (NN) Pb-Pb distance $d = 3.00$ \AA \ are obtained through structural optimization, which agree with the previous calculations\cite{NC4.1500}.
The space group is $P\bar{3}m1$.
The angle $\phi$, defined as being between the Pb-Pb bond and the $z$ direction in Fig. \ref{fig:LattiStruct} (c), is $108.34 ^ \circ$. In comparison to graphene with $\phi = 90 ^ \circ$, the larger Pb-Pb bond length weakens the $\pi - \pi$ overlap. Meanwhile, the buckling can enhance the overlap between $\sigma$ and $\pi$ orbitals and stabilizes the system. The same mechanism also applies for other 2D materials in group IV which share the similar buckled configuration.

\subsection{First-principles calculations}
In this paper, our first-principles calculations for electronic structures of plumbene were carried out using the WIEN2K package with a full-potential augmented plane wave\cite{wien2k}, based on the Perdew-Burke-Ernzerhof generalized gradient approximation (PBE-GGA)\cite{PRL77.3865}. Besides, the energy gaps of plumbene have also been investigated with the local density approximation (LDA)\cite{PRB45.13244}, the Wu-Cohen GGA (WC-GGA)\cite{PRB73.235116} and a combination of the modified Becke-Johnson exchange potential and the PBE-GGA correlation potential (mBJ-PBE)\cite{PRL102.226401}. The number of $k$ points is $10000$ in the Brillouin zone. The muffin-tin radius is $2.5$ Bohr radius for Pb and $R_{mt}K_{max}$ is chosen to be $8.5$. SOC is included in all the first-principles calculations unless stated otherwise.

\subsection{Tight-binding model}
In most of the previous works about $2$D group-IV materials, only the NN hopping is considered in tight-binding models, because the itineracy of electrons is bad and they focus on the band structure near the Fermi level\cite{PRB84.195430,NC4.1500,JPCC119.11896,PRB82.245412,NSPI978}. There is no definitive strategy to fit the bands. In this study, both the NN and next-nearest-neighbor (NNN) hoppings are introduced into the total Hamiltonian so that the two bands above $E_{F}$ can be fitted better.
Three NN and six NNN translation vectors are
\begin{equation}
\begin{array}{l}
\mathbf {{d_1}}  = \frac{a}{{\sqrt 3 }}\left( {\begin{array}{*{20}{c}}
{\frac{{\sqrt 3 }}{2}}&{\frac{1}{2}}&{\cot \phi }
\end{array}} \right),\;\\
\mathbf {{d_2}}  = \frac{a}{{\sqrt 3 }}\left( {\begin{array}{*{20}{c}}
{ - \frac{{\sqrt 3 }}{2}}&{\frac{1}{2}}&{\cot \phi }
\end{array}} \right),\;\\
\mathbf {{d_3}}  = \frac{a}{{\sqrt 3 }}\left( {\begin{array}{*{20}{c}}
0&{ - 1}&{\cot \phi }
\end{array}} \right),
\end{array}
\end{equation}
\begin{equation}
\begin{array}{l}
\mathbf {{v_1}}  = a\left( {\begin{array}{*{20}{c}}
{\frac{1}{2}}&{\frac{{\sqrt 3 }}{2}}&0
\end{array}} \right),\;\mathbf {{v_2}}  = a\left( {\begin{array}{*{20}{c}}
1&0&0
\end{array}} \right),\;\\
\mathbf {{v_3}}  = a\left( {\begin{array}{*{20}{c}}
{\frac{1}{2}}&{ - \frac{{\sqrt 3 }}{2}}&0
\end{array}} \right),\;
\mathbf {{v_4}}  = a\left( {\begin{array}{*{20}{c}}
{ - \frac{1}{2}}&{ - \frac{{\sqrt 3 }}{2}}&0
\end{array}} \right),\;\\
\mathbf {{v_5}}  = a\left( {\begin{array}{*{20}{c}}
{ - 1}&0&0
\end{array}} \right),\;\mathbf {{v_6}}  = a\left( {\begin{array}{*{20}{c}}
{ - \frac{1}{2}}&{\frac{{\sqrt 3 }}{2}}&0
\end{array}} \right),
\end{array}
\end{equation}
where $a$ is defined as the NNN distance.
The total Hamiltonian read
\begin{equation}
H = {H_{00}} + {H_0} + H_{A0}^{'} + H_{B0}^{'} + {H_{so}},
\label{equ:Htot}
\end{equation}
where the five terms correspond to on-site energy, NN hopping, NNN hopping in $A$ and $B$ sublattices, and effective on-site SOC interaction, respectively. They reads
\begin{equation}
{H_{00}} = \sum\limits_{L, i, \alpha, \sigma } {{\varepsilon _\alpha }c_{L i\alpha \sigma }^\dag {c_{L i\alpha \sigma }}} ,
\end{equation}
\begin{equation}
\begin{array}{l}
{H_0} = \sum\limits_{i,\delta ,\alpha ,\alpha ',\sigma } {{t_{AB\delta \alpha \alpha '}}c_{A,i + \delta ,\alpha \sigma }^\dag {c_{Bi\alpha '\sigma }}} \\
 + \sum\limits_{i,\eta ,\alpha ,\alpha ',\sigma } {{t_{BA\eta \alpha '\alpha }}c_{B,i + \eta ,\alpha '\sigma }^\dag {c_{Ai\alpha \sigma }}},
\end{array}
\end{equation}
\begin{equation}
H_{A0}^{'} = \sum\limits_{i,v,\sigma } {{t_{AAv\alpha \alpha '}}c_{A,i + v,\alpha \sigma }^\dag {c_{Ai\alpha '\sigma }}},
\label{equ:HA0p}
\end{equation}
\begin{equation}
{H_{so}} = \sum\limits_i {\left( \begin{array}{l}
i\frac{{{\xi _0}}}{2}c_{Aiz \uparrow }^\dag {c_{Aiy \downarrow }} - i\frac{{{\xi _0}}}{2}c_{Aiy \downarrow }^\dag {c_{Aiz \uparrow }} + i\frac{{{\xi _0}}}{2}c_{Aiz \downarrow }^\dag {c_{Aiy \uparrow }}\\
 - i\frac{{{\xi _0}}}{2}c_{Aiy \uparrow }^\dag {c_{Aiz \downarrow }} + i\frac{{{\xi _0}}}{2}c_{Aiy \uparrow }^\dag {c_{Aix \uparrow }} - i\frac{{{\xi _0}}}{2}c_{Aix \uparrow }^\dag {c_{Aiy \uparrow }}\\
 - i\frac{{{\xi _0}}}{2}c_{Aiy \downarrow }^\dag {c_{Aix \downarrow }} + i\frac{{{\xi _0}}}{2}c_{Aix \downarrow }^\dag {c_{Aiy \downarrow }} - \frac{{{\xi _0}}}{2}c_{Aiz \uparrow }^\dag {c_{Aix \downarrow }}\\
 - \frac{{{\xi _0}}}{2}c_{Aix \downarrow }^\dag {c_{Aiz \uparrow }} + \frac{{{\xi _0}}}{2}c_{Aiz \downarrow }^\dag {c_{Aix \uparrow }} + \frac{{{\xi _0}}}{2}c_{Aix \uparrow }^\dag {c_{Aiz \downarrow }}\\
 + i\frac{{{\xi _0}}}{2}c_{Biz \uparrow }^\dag {c_{Biy \downarrow }} - i\frac{{{\xi _0}}}{2}c_{Biy \downarrow }^\dag {c_{Biz \uparrow }} + i\frac{{{\xi _0}}}{2}c_{Biz \downarrow }^\dag {c_{Biy \uparrow }}\\
 - i\frac{{{\xi _0}}}{2}c_{Biy \uparrow }^\dag {c_{Biz \downarrow }} + i\frac{{{\xi _0}}}{2}c_{Biy \uparrow }^\dag {c_{Bix \uparrow }} - i\frac{{{\xi _0}}}{2}c_{Bix \uparrow }^\dag {c_{Biy \uparrow }}\\
 - i\frac{{{\xi _0}}}{2}c_{Biy \downarrow }^\dag {c_{Bix \downarrow }} + i\frac{{{\xi _0}}}{2}c_{Bix \downarrow }^\dag {c_{Biy \downarrow }} - \frac{{{\xi _0}}}{2}c_{Biz \uparrow }^\dag {c_{Bix \downarrow }}\\
 - \frac{{{\xi _0}}}{2}c_{Bix \downarrow }^\dag {c_{Biz \uparrow }} + \frac{{{\xi _0}}}{2}c_{Biz \downarrow }^\dag {c_{Bix \uparrow }} + \frac{{{\xi _0}}}{2}c_{Bix \uparrow }^\dag {c_{Biz \downarrow }}
\end{array} \right)}
\end{equation}
where $c_{Li\alpha \sigma }^\dag $ and ${c_{Li\alpha \sigma }}$ represent the electron creation and annihilation operators at site $i$ of sublattice $L$ for electrons with spin $\sigma $ and orbital $\alpha $, respectively. $\delta$ and $\eta$ correspond to NN vectors of the honeycomb lattice and $\eta=-\delta$. ${t_{AB\delta \alpha \alpha '}}$ is the hopping integral along $\mathbf \delta $ from orbital $\alpha '$ of sublattice $B$ to orbital $\alpha$ of sublattice $A$ and is given by the Slater-Koster formula\cite{PR94.1498}. In Table \ref{tab:param} the eight Slater-Koster parameters ${V_{ss\sigma }},{V_{sp\sigma }},{V_{pp\sigma }},{V_{pp\pi}},V{'_{ss\sigma }},V{'_{sp\sigma }},V{'_{pp\sigma }},V{'_{pp\pi }}$ correspond to the $\sigma$ and $\pi$ bonds formed by $6s$ and $6p$ orbitals of NN and NNN sites. The other three fitted parameters ${\varepsilon _s}$, ${\varepsilon _p}$ and ${\xi _0}$ are the energy levels of $s$ and $p$ orbitals and effective on-site SOC strength, respectively. For ${H_{SO}}$ the effective on-site SOC coefficients among atomic orbitals are determined by following earlier theoretical studies\cite{PRB84.195430}. $H_{B0}^{'}$ can be obtained by replacing the index $A$ with $B$ in Eq. (\ref{equ:HA0p}).

\begin{table}[!hbp]
\caption{The NN and NNN hopping integrals between $s$ and $p$ orbitals are respectively listed in the left and right columns and are considered as functions of the direction cosine $l$ ($l'$), $m$ ($m'$) and $n$ ($n'$) of the vector $\mathbf d $ ($\mathbf v$). Other hopping integrals are found by permuting indices.}
\begin{tabular}{cccccccc}
\hline
\hline
\ \ & \ \ \ NN \ \ & \ \ \  NNN \\
\hline
$t_{ss}$ \ \ & \ \ \ $V_{ss\sigma}$ \ \ & \ \ \ $V{'_{ss\sigma }}$ \\
$t_{sx}$ \ \ & \ \ \ $lV_{sp\sigma}$ \ \ & \ \ \ $l'V{'_{sp\sigma }}$ \\
$t_{xs}$ \ \ & \ \ \ $-lV_{sp\sigma}$ \ \ & \ \ \ $-l'V{'_{sp\sigma }}$ \\
$t_{xx}$ \ \ & \ \ \ $l^{2}V_{pp\sigma}+(1-l^{2})V_{pp\pi }$ \ \ & \ \ \ $l'^{2}V'_{pp\sigma}+(1-l'^{2})V'_{pp\pi }$ \\
$t_{xy}$ \ \ & \ \ \ $lm(V_{pp\sigma}-V_{pp\pi})$ \ \ & \ \ \ $l'm'(V'_{pp\sigma}-V'_{pp\pi})$ \\
$t_{yz}$ \ \ & \ \ \ $mn(V_{pp\sigma}-V_{pp\pi})$ \ \ & \ \ \ $m'n'(V'_{pp\sigma}-V'_{pp\pi})$ \\
\hline
\hline
\label{tab:param}
\end{tabular}
\end{table}

When external perpendicular electric and magnetic fields ($E_z$ and $B_z$) are applied, the corresponding Hamiltonians can be respectively expressed as follows,
\begin{equation}
\begin{array}{l}
{H_E} = {E_z}\sum\limits_{m,i,\alpha ,\sigma } {{z_m}c_{m i\alpha \sigma }^ + {c_{m i\alpha \sigma }}} ,\\
\end{array}
\end{equation}
\begin{equation}
\begin{array}{l}
{H_B} =  - {B_z}\sum\limits_{m,i,\alpha } {\left( {c_{m i\alpha  \uparrow }^ + {c_{m i\alpha  \uparrow }} - c_{m i\alpha  \downarrow }^ + {c_{m i\alpha  \downarrow }}} \right)},\\
\end{array}
\end{equation}
where $z_m$ is the $z$ coordinate of the sublattice $m$.

\section{Results and discussion}
\subsection{Electronic structures $\And$ topological characteristics}
\begin{figure}[tbp]
\centering
\setlength{\abovecaptionskip}{2pt}
\setlength{\belowcaptionskip}{4pt}
\includegraphics[angle=0, width=8cm]{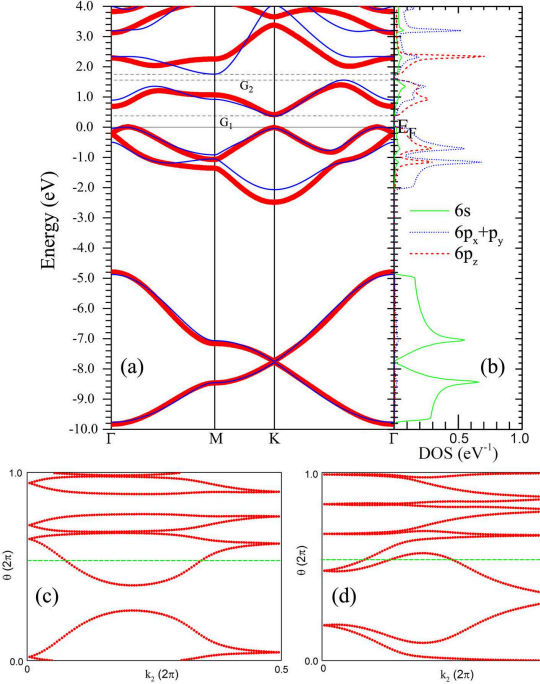}
  \caption{(Color online) (a) Band structures of plumbene obtained from the first-principles (thin blue lines) and tight-binding (thick red lines) calculations. Two gaps around $E = 0$ and $1.7$ eV are marked by $G_1$ and $G_2$, respectively. (b) Density of states projected into Pb-$6s$ and $6p$ orbitals. Wannier centers with (c) $E_F$ inside the gap $G_1$ (the electron-doping concentration $n=0$) and (d) $E_F$ inside the gap $G_2$ ($n=2$). The red dotted and green dashed lines are the evolution and reference lines, respectively.}
  \label{fig:Bandso-WanC}
\end{figure}

Fig. \ref{fig:Bandso-WanC} (a) and (b) show the band structure and density of states (DOS) of plumbene from first-principles calculations, respectively. The states around the Fermi level are dominantly contributed by Pb-$6p$ orbitals.
Three isolated bands are formed near the Fermi level and an energy gap of $\sim 400$meV is opened at $E_F$. The gap is much larger than that of the other group-IV $2$D materials, which are $10^{-3}, 8, 23, 72$ meV from theoretical calculations for graphene, silicene, germanene and stanene, respectively\cite{PRB75.041401,NJP16.105018,JPCC119.11896}. Although plumbene has stronger SOC interaction, leading to a larger energy gap, it is totally different from the other four topologically nontrivial systems and exhibits a normal insulating behavior\cite{SR6.20152,NJP16.105018}. Besides the gap at $E_{F}$, we find another gap ($\sim 200$ meV) between the conduction band and higher energy bands. The two gaps are denoted as $G_1$ and $G_2$ in Fig. \ref{fig:Bandso-WanC} (a).

\begin{center}
\linespread{1.5}
\begin{table}[!hbp]
\caption{The parameters are fitted to the first-principles calculations in the tight-binding model in units of eV.}
\begin{tabular}{cccccccc}
\hline
\hline
          $\varepsilon _s$ \ \
 & \ \ \  $\varepsilon _p$  \ \
 & \ \ \ $V_{ss\sigma }$  \ \
 & \ \ \ $V_{sp\sigma }$  \ \
 & \ \ \ $V_{pp\sigma }$  \ \
 & \ \ \ $V_{pp\pi }$ \\
\hline
          $-7.009$ \ \
 &  \ \ \ $1.388$  \ \
 &  \ \ \ $- 0.879$ \ \
 &  \ \ \ $1.326$ \ \
 &  \ \ \ $1.990$ \ \
 &  \ \ \ $- 0.676$ \\
\hline
\hline
          $V{'_{ss\sigma }}$ \ \
 & \ \ \  $V{'_{sp\sigma }}$  \ \
 & \ \ \ $V{'_{pp\sigma }}$  \ \
 & \ \ \ $V{'_{pp\pi }}$  \ \
 & \ \ \ ${\xi _0}$   \\
\hline
          $- 0.015$ \ \
 &  \ \ \ $- 0.109$  \ \
 &  \ \ \ $0.253$ \ \
 &  \ \ \ $- 0.103$ \ \
 &  \ \ \ $0.800$ \\
\hline
\hline
\label{tab:fitting}
\end{tabular}
\end{table}
\end{center}

In order to study the properties of plumbene in detail, including its topological characteristics and microscopic responses to external fields, we have employed a sixteen-band tight-binding model with NN and NNN hoppings to fit the band structure according to the first-principles result (see Table \ref{tab:fitting}).
The corresponding band structures are shown in Fig. \ref{fig:Bandso-WanC} (a).
By comparing the model bands with first-principles bands, we can see that they agree well with each other in the energy range ($-10$ eV $\sim 3$ eV).

To determine whether plumbene is topologically trivial or nontrivial, we have studied the Wannier center. Its evolution with $k_2$ can be easily obtained by looking at the phase factor $\theta \left( {{k_2}} \right)$\cite{PRB84.075119}, which is an equivalent expression for the $Z_2$ topological invariant. $k_2$ is defined as
${k_2} = \mathbf {{k_2}}  \cdot \mathbf {{a_2}} .\;\mathbf {{k_2}}  = \left( {\begin{array}{*{20}{c}}
0&0
\end{array}} \right)\sim\left( {\begin{array}{*{20}{c}}
{ - \frac{\pi }{a_{0}}}&{\frac{\pi }{{\sqrt 3 a_{0}}}}
\end{array}} \right)$
along $\mathbf {{b_2}}$ direction. $\mathbf {{a_2}}$ and $\mathbf {{b_2}}$ are defined in Fig. \ref{fig:LattiStruct}.
The Wannier centers are calculated based on the tight-binding model and shown in Fig. \ref{fig:Bandso-WanC} (c) and (d). When the Fermi level is inside the gap $G_1$, through moving the green reference line we find it always crosses the evolution lines of Wannier centers zero or two (even) times, indicating that the system is topologically trivial. In contrast, for the case that the Fermi level is inside the gap $G_2$, the reference line always crosses the evolution lines one or three (odd) times, indicating that the system is topologically nontrivial.

\begin{figure}[tbp]
  \centering \includegraphics[width=8.0cm]{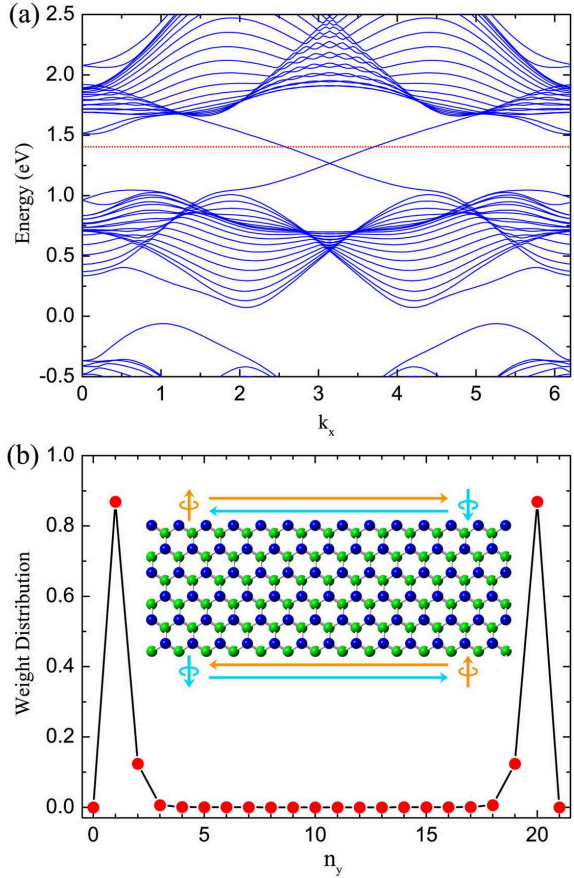}
  \caption{(Color online) (a) The band structure of the zigzag plumbene nanoribbon with a width $N_{y} = 20$. (b) The spacial weight distributions corresponding to the states that are on the red dotted line in (a). The states inside the gap $G_2$ are edge states. Inset: Schematic of the spin-resolved edge channels in the plumbene nanoribbon.}
  \label{fig:ZigzBand}
\end{figure}
Topological insulators are characterized by topologically protected metallic edge states with helical spin polarization residing inside an insulating bulk gap. To determine whether plumbene with electron doping can be topologically nontrivial, we have studied plumbene nanoribbon. The width of nanoribbon is selected to be large enough to avoid interactions between edge states.
Without loss of generality, we consider a zigzag nanoribbon, which is periodic along $x$ direction with open boundaries in $y$ direction. There are $N_y$ zigzag chains.
Fig. \ref{fig:ZigzBand} (a) shows the band structure of the zigzag plumbene ribbon with $N_{y} = 20$. Through comparing it with the bulk band structure, we find that two bands cross the bulk gap $G_2$ in the ribbon model. Through a weight analysis, we can determine that the states within the bulk gap $G_2$ are edge states, as shown in Fig. \ref{fig:ZigzBand} (b). The largest contribution are from the zigzag chains at the boundaries, including $n_{y}=1$,$2$,$19$ and $20$. The weight at inner sites can be negligible. In contrast, the gap $G_1$ is fully opened. It corresponds to a normal band insulator, which is in agreement with previous studies\cite{SR6.20152,NJP16.105018}. Through further spin-resolved analysis, the edge states within the bulk gap $G_2$ show a distinct helical property: two states with opposite spin-polarization counterpropagate at each edge, as illustrated in the inset of Fig. \ref{fig:ZigzBand} (b).
We have also test other cases with different $N_y$ ($e.g.$ $10$,$30$,$50$) and get qualitatively consistent results. In addition, the ribbon with armchair-type edges has also been studied and similar results are obtained.
Here, we report the nontrivial topological characteristic of plumbene without any chemical functionalization. Such a large bulk gap also make plumbene be an ideal candidate for realizing the QSH effect at room temperature.

\begin{figure}[tbp]
\centering
\setlength{\abovecaptionskip}{2pt}
\setlength{\belowcaptionskip}{4pt}
\includegraphics[angle=0, width=1.0 \columnwidth]{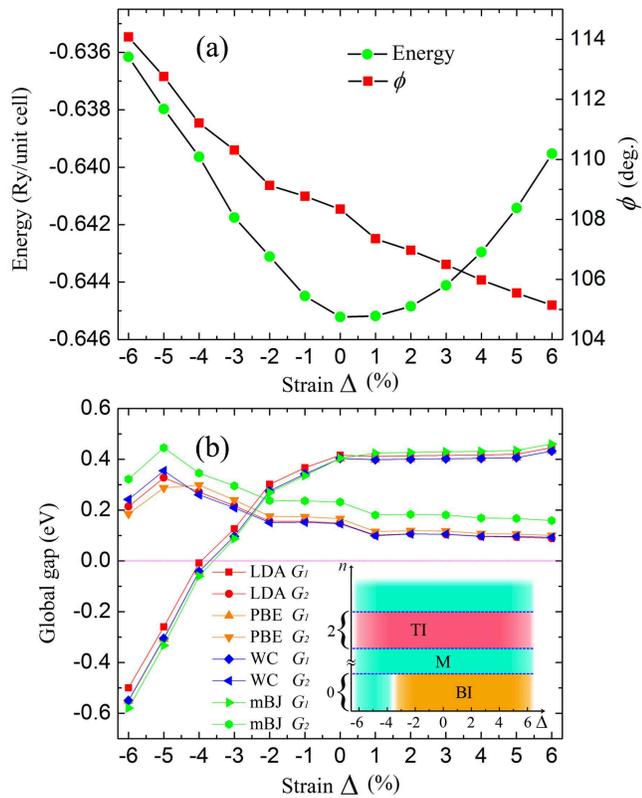}
  \caption{(Color online) (a) Total energy per unit cell and $\phi$ as functions of $\Delta$. (b) Two gaps $G_{1}$ and $G_{2}$ as functions of $\Delta$ calculated with LDA, PBE-GGA, WC-GGA and mBJ-PBE exchange-correlation functionals. Inset: The $\Delta - n$ phase diagram of plumbene. The horizontal and vertical axes correspond to the strain $\Delta$ and the electron doping concentration $n$. The red, orange and green regions represent topological insulating (TI), band insulating (BI) and metallic (M) phases, respectively.}
  \label{fig:Ene-gap-PhD}
\end{figure}

To show the robustness of the topological properties in plumbene, we have also studied adiabatic evolution of the electronic structures under external in-plane strain $\Delta$, which is defined as $\Delta  = {{\left( {a - {a_0}} \right)} \mathord{\left/
{\vphantom {{\left( {a - {a_0}} \right)} {{a_0}}}} \right.  \kern-\nulldelimiterspace} {{a_0}}} \times 100\%$. The $z$ coordinates of Pb atoms are fully relaxed.
In the range of $- 6\%  < \Delta  < 6\%$, the energy bands near the Fermi level are always in isolation, as plotted in Fig. S1 of the Supplemental Material\cite{supplementary}. So the topology of the system remains unchanged.
Fig. \ref{fig:Ene-gap-PhD} (a) shows the evolutions of the total energy and the angle $\phi$ under different strains. The structure with $\Delta = 0$, corresponding to the optimized lattice structure, is the lowest-energy one. $\phi$ decreases monotonously with increasing $\Delta$.
The gaps $G_{1}$ and $G_{2}$ are investigated with four approximations to the exchange-correlation functionals, including LDA\cite{PRB45.13244}, PBE-GGA\cite{PRL77.3865}, WC-GGA\cite{PRB73.235116} and mBJ-PBE\cite{PRL102.226401}.
For energy gaps of $sp$-semiconductors, the mBJ potential performs in general better than others\cite{wien2k}. Fig. \ref{fig:Ene-gap-PhD} (b) shows that the gap $G_2$ by mBJ-PBE is about $100$ meV larger than  that by the other three approximations. On the whole, their results are in qualitative agreement with each other.
The two gaps versus $\Delta$ show totally different behaviors. $G_1$ increases with increasing $\Delta$, while $G_2$ remains relatively stable. When the compressive strain is excessively large, the system is transformed into a metallic phase due to two bands crossing the Fermi level. It also leads to a negative gap (Supplementary Fig. S1 (a)\cite{supplementary}). With increasing the value of $\Delta$, the global gap $G_{1}$ changes from an indirect gap to a direct gap, while the global gap $G_{2}$ always keeps an indirect gap.
The Fermi level can be shifted to the top of the conduction band using gating methods by electric fields\cite{APL90.052905,NN10.270} and the corresponding doping concentration $n = 2$. For this doping level, when the strain $\Delta$ changes from $-6 \%$ to $6 \%$, there are not any band crossings near the Fermi level and the system keeps topologically nontrivial. So the topological insulator state of plumbene is a robust state with respect to a certain range of in-plane strain. Its indirect bulk gap also enhances the stability. Based on the above results, we can obtain a $\Delta  - n$ phase diagram, as plotted in the inset of Fig. \ref{fig:Ene-gap-PhD} (b). And both the large gaps $G_1$ and $G_2$ are ample for applications at room temperature.

\subsection{External-field responses $\And$ device design}
\begin{figure}[tbp]
\centering
\setlength{\abovecaptionskip}{2pt}
\setlength{\belowcaptionskip}{4pt}
\includegraphics[angle=0, width=1.0 \columnwidth]{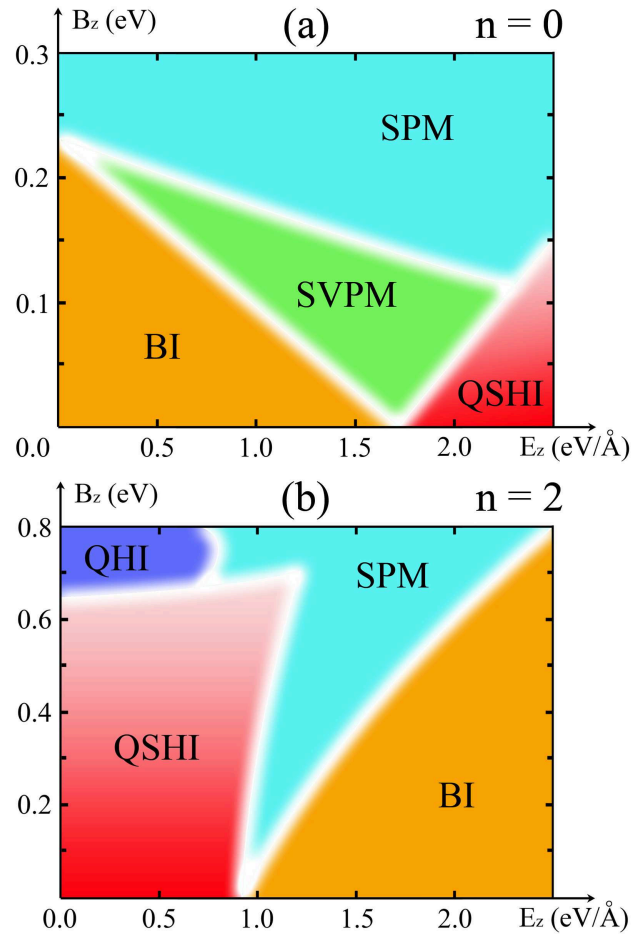}
  \caption{(Color online) (a, b) The $E_{z} - B_{z}$ phase diagram of plumbene without and with electron doping $n=2$, based on the tight-binding model. The orange, green, blue, red and purple regions represent band insulator (BI), spin valley polarized metal (SVPM), spin polarized metal (SPM), quantum spin Hall insulator (QSHI) and quantum Hall insulator (QHI), respectively. In a strict sense, only the QSHI region on the $E_z$ axis is robust; the edge states are unstable for $B_z>0$.}
  \label{fig:PhaseDiag-E-B}
\end{figure}

Plumbene, analogous to graphene\cite{PRL97.016801,PRL99.236809}, has extraordinary electronic properties. There are two degenerate and inequivalent valleys at the corners of the Brillouin zone ($K$ and $K'$ points). Intervalley scattering is strongly suppressed due to their large separation in momentum space. Moreover, because of the buckled structure, the presence of an external perpendicular electric field can induce a voltage difference between the two sublattices.
An electric field $E_z$ is assumed to be homogeneous along the $z$ direction in the tight-binding model and it can lead to a high spin polarization at $K$ and $K'$ points (see the Supplemental Material for details\cite{supplementary}).
Even so, the system remains nonmagnetic due to their degeneracy.
In order to realize the spin-polarized transport, we further consider an external perpendicular magnetic field $B_z$.
When both electric and magnetic fields are applied, the degeneracy of bands is almost lifted in the entire Brillouin zone, including at the time-reversal invariant points.
With increasing the strength of fields, the band splitting is further increased.
Fig. \ref{fig:PhaseDiag-E-B} (a) shows an $E_{z} - B_{z}$ phase diagram in a pure plumbene ($n = 0$). When the external fields are small, the system is a normal band insulator. With increasing the fields, the spin is polarized gradually and the degeneracy between $K$ and $K'$ points is also lifted. When one of the two valleys is occupied by electrons, the system becomes a spin valley polarized metal. It is worth mentioning that the electrons in the occupied valley ($e.g.$ $K'$ point) are from the states around $\Gamma$ and $K$ points in the spin valley polarized metallic phase, while for graphene and silicene the electrons in the occupied valley are only from the latter. With excessively large electric and magnetic fields, the valley polarization vanishes, resulting in a spin polarized metal. In the absence of $B_{z}$, the increasing $E_{z}$ makes the gap $G_1$ close and reopen. Simultaneously, the system changes from a band insulator to a QSH insulator.
Furthermore, we have also investigated the response of plumbene to the external fields with electron doping $n=2$. The corresponding $E_{z} - B_{z}$ phase diagram is plotted in Fig. \ref{fig:PhaseDiag-E-B} (b). In small external fields the system remains topologically nontrivial as a QSH insulator. With increasing $B_{z}$, the spin of the edge states is polarized, leading to a quantum Hall insulator. For a small magnetic field, the increasing electric field can destroy the edge states and the system is transformed into a band insulator. Similar to the case without electron doping, when both $E_{z}$ and $B_{z}$ are excessively large, the system becomes a spin polarized metal. Since the above results are based on the tight-binding model fitted to the first-principles band structure, the two phase diagrams are qualitatively correct for describing quantum phase transitions.

\begin{figure}[tbp]
\centering
\setlength{\abovecaptionskip}{2pt}
\setlength{\belowcaptionskip}{4pt}
\includegraphics[angle=0, width=8cm]{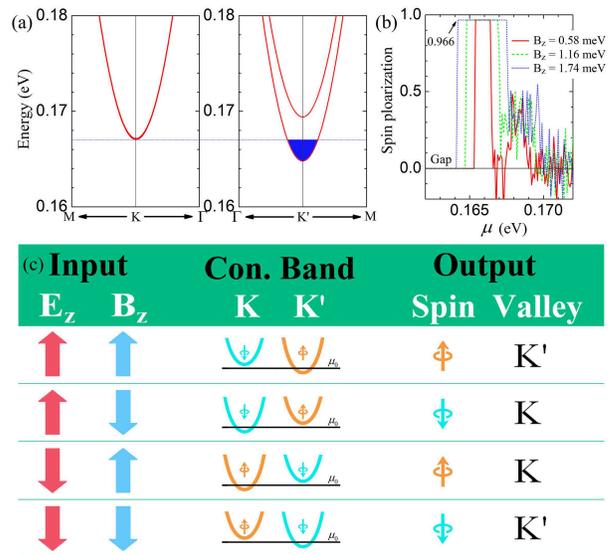}
  \caption{(Color online) (a) Conduction bands near $K$ and $K'$ points with $E_z = 10$ meV/\AA, $B_z = 1.16$ meV. (b) Spin polarizations as functions of chemical potential $\mu$ with $E_z = 10$ meV/\AA \ and $B_z = 0.58, 1.16, 1.74$ meV. The horizontal axis corresponds to the vertical axis in (a). (c) Schematic diagram of a four-state spin-valley filter. Left column: four combinations of $E_z$ and $B_z$ with different directions as inputs. Middle column: spin-split conduction bands at $K$ and $K'$ points. The chemical potential is tuned to $\mu_0$. Right column: four spin-valley electric states as outputs. }
  \label{fig:filter}
\end{figure}

Fig. \ref{fig:filter} (a) shows the conduction bands near $K$ and $K'$ points with $E_z = 10$ meV/\AA \ and $B_z = 1.16$ meV, which respectively correspond to the electric field intensity $100$ MV/m and the magnetic induction intensity $20$ T. The degeneracy between two valleys is lifted and the valley at $K'$ points is lower than that at $K$ point. The corresponding spin polarization remains large and reaches $96.6 \%$, as shown in Fig. \ref{fig:filter} (b). Moreover, with increasing $B_z$ the spin polarization at the bottom of the conduction band almost remains constant but the energy range of the polarization increases. The carrier concentration can be controlled and tuned into the blue area in the band structure, and then the spin and valley polarizations can be realized simultaneously without other bands crossing the Fermi level. Therefore, plumbene with the perpendicular electric and magnetic fields can serve as a spin-valley filter, with spin sign and valley switchable by altering the field directions.

Following this idea, we present a schematic diagram of a four-state spin-valley filter in Fig. \ref{fig:filter} (c). For an experimental realization of the proposal, there are three preliminary steps to be made. (i) External perpendicular fields $E_z$ and $B_z$ are applied on plumbene as a controller. (ii) The chemical potential is tuned to the energy $\mu_0$ which is marked in the middle column of Fig. \ref{fig:filter} (c). This progress can be feasible using gating methods in experiments\cite{APL90.052905,NN10.270}. Now the spin and valley polarizations are achieved simultaneously. (iii) Through fixing the strengths of the external fields and just changing their directions we can obtain four different output states. This device can be of utility as a four-state spin-valley filter, which can be applied in spintronics, valleytronics and even the hybrid structures of them.

Besides, there exist topologically protected edge states in the bulk gap $G_2$. As is well known, the edge states of topological insulators with strong SOC interaction are usually regarded as a spin generator\cite{NM11.409}. When the position of the Fermi level is tuned into the gap $G_2$, plumbene can become a spin generator with a large bulk gap, which can help to screen the interference of thermally activated carriers and allow a large controlling range of external fields. Although Bi monolayer film has also been predicted as a topological insulator with a larger bulk gap, there are no conduction-band valleys at $K$ and $K'$ points\cite{PRL107.136805,PRL97.236805,PRB83.121310,SR5.11512,NJP16.105018}, so that Bi monolayer film can not be used to fabricate the four-state spin-valley filter mentioned above.

\subsection{Substrate effects}
Normally, in experiments a $2$D film needs to be suspended on a certain substrate which may induce some strain on the film due to lattice mismatch and affect the electronic structure. The study of the electronic structures under different in-plane strains has shown the robustness of topological properties in plumbene. This means that it is insensitive to the in-plane strain of the substrate. In reality, the substrate effect is not usually restricted to stretching or compression of the $2$D film by the substrate. The interaction between them, such as chemical bonding, may strongly modify the electronic structure and properties\cite{NP6.104,NC1.17,NM14.1020}. Assuming that plumbene can be synthesized experimentally, there are several suggestions for future experiments about the substrate effect. (i) Choose the substrate which induces as little effect as possible on the electronic structure of the $2$D Pb film. Specially, it is better to avoid the bonding between Pb atoms and the substrate. (ii) When the band structure near the Fermi level is modified slightly, we can adjust it by the perpendicular electric field, based on the responses of plumbene to external fields. (iii) If the properties of the film are changed dramatically (\textit{e. g.}, the system becomes a metal), one can perform a chemical functionalization with appropriate adsorbates (such as CH$_3$ that the corresponding functionalization was observed to show much more moderate reaction kinetics than plasma fluorination and hydrogenation\cite{NC5.3389,NanoL15.1083}), in order to reduce the substrate effect and keep the global gap open. But the topological characteristic may be changed, which depends on the chemical functionalization. Meanwhile, both the valleys at $K$ and $K'$ points may vanish due to the saturation of $p_z$ dangling bonds.

Recently Zhou \textit{et al.} theoretically showed that the Si$(111)$ surface functionalized with one-third monolayer of hydrogen or halogen atoms exhibiting a trigonal superstructure provides an ideal template for epitaxial growth of heavy metals, such as Bi, Pb and Tl, which self-assemble into a hexagonal lattice with high kinetic and thermodynamic stability\cite{PNAS111.14378,SR4.7102}. The template of a functionalized Si surface has already been widely studied and used in early surface science research\cite{CR102.1271,SS163.457,JVSTA23.1100,JPCC113.21713,NM9.266,Science268.1590,NC4.1649}. Most remarkably, the hexagonal metal overlayer is atomically bonded to but electronically decoupled from the underlying Si substrate, exhibiting large-gap QSH states completely isolated from Si valence and conduction bands in Bi@X-Si$(111)$ ( X = H, Cl, Br, I) and Pb@H-Si$(111)$. Moreover, hexagonal lattices of indium overlayer have been successfully grown on the Si$(111) - \sqrt{3} \times \sqrt{3} - $Au surface in experiments\cite{PRB73.115335}. Therefore, it is highly feasible to experimentally realize the large-gap QSH states in the hexagonal Pb film, based on the existing related experiments and theoretical calculations. We also note that the substrate may hence the topological properties of plumbene, which are of interest and will be explored in our future studies.

\section{Summary}
In summary, we employed first-principles and tight-binding methods to investigate the electronic structures and topological properties of plumbene. Two large bulk gaps were predicted. Both of them allow for viable applications at room temperature. The $n - \phi$ and $E_{z} - B_{z}$ phase diagrams show novel and rich phase transitions. Moreover, we find that the QSH state is very robust with respect to external in-plane strain. The large bulk gap and stable edge states make plumbene an ideal candidate for realizing the QSH effect at room temperature. According to the responses of plumbene in external perpendicular electric and magnetic fields, we propose a design for a four-state spin-valley filter.
The substrate effects of plumbene have also been discussed.
Plumbene may open a new and exciting avenue for realizing the QSH effect and designing spintronic devices at room temperature.

\section*{Acknowledgments}
This work is supported by the National Natural Science Foundation of China (Grants No. 11681240276, No. 11674152 and No. 11604142), and the Shenzhen Peacock Plan and Shenzhen Fundamental Research Foundation (Grant No. JCYJ20150630145302225).

\end{document}


\title{Supplemental Material for ``From a normal insulator to a topological insulator in plumbene''}
\author{Xiang-Long Yu}
\affiliation{Department of Physics, South University of Science and Technology of China, Shenzhen 518055, P.R. China}
\affiliation{School of Physics and Technology, Wuhan University, Wuhan 430072, P.R. China}
\author{Li Huang}
\affiliation{Department of Physics, South University of Science and Technology of China, Shenzhen 518055, P.R. China}
\author{Jiansheng Wu}
\email{wujs@sustc.edu.cn}
\affiliation{Department of Physics, South University of Science and Technology of China, Shenzhen 518055, P.R. China}

\maketitle

\begin{figure}[tbp]
  \centering \includegraphics[width=16.0cm]{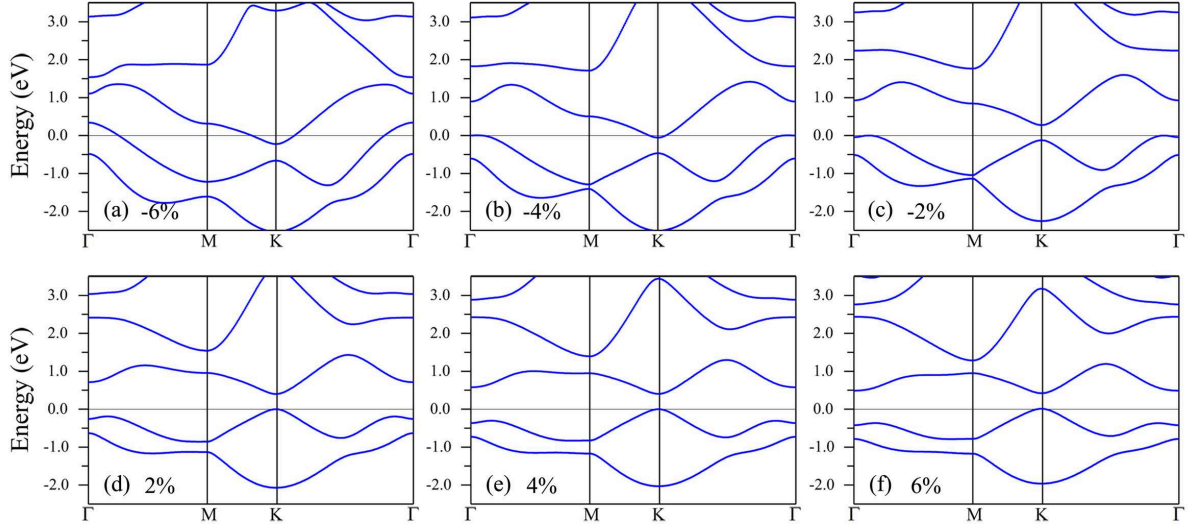}
  \renewcommand{\figurename}{Fig. S}
  \caption{(Color online) (a-f) The band structures of plumbene with strain $\Delta = -6\%,$ $-4\%,$ $-2\%,$ $2\%,$ $4\%$ and  $6\%$ from first-principles calculations.}
  \label{fig:Band-Strain}
\end{figure}
Fig. S\ref{fig:Band-Strain} shows the evolution of bands near the Fermi level under different in-plane strains, and the result  corresponds to Fig. 4 in the main text.

\begin{figure}[tbp]
  \centering \includegraphics[width=15.0cm]{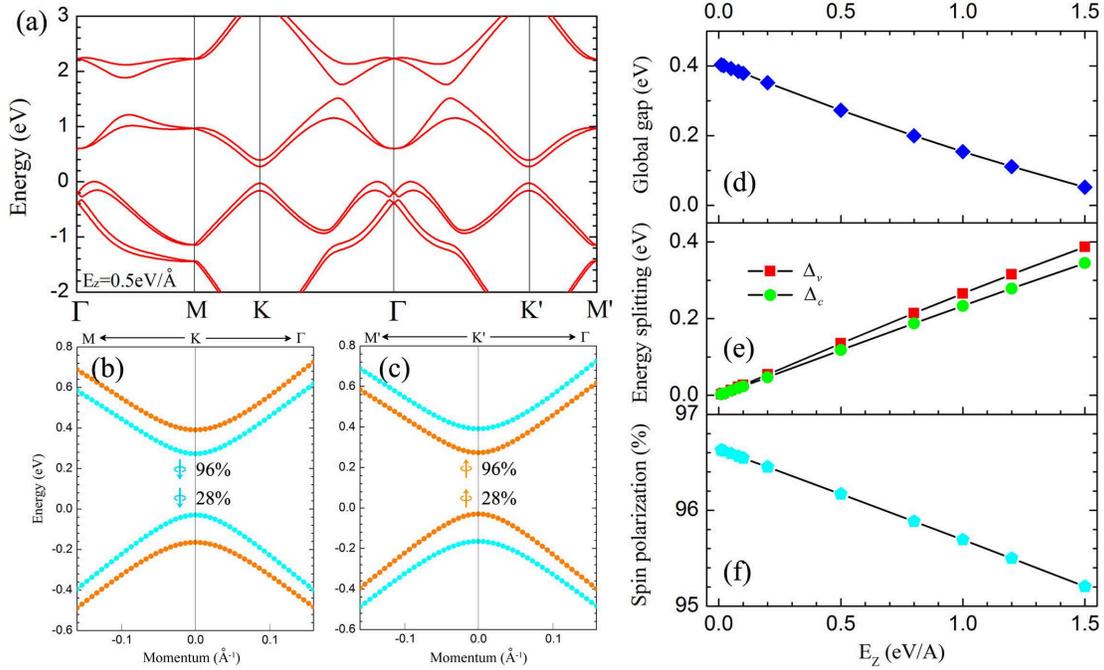}
  \renewcommand{\figurename}{Fig. S}
  \caption{(Color online) (a-c) The band structures of plumbene in the presence of an perpendicular electric field $E_{z}=0.5$ eV/{\AA}. The color and the arrow represent the spin-polarized direction.
  (d) The global band gap, (e) the energy splittings of the conduction ($\Delta_c$) and valence ($\Delta_v$) bands at $K$ point and (f) the spin polarization ratio of the conduction band at $K$ point as functions of the perpendicular electric field $E_z$.}
  \label{fig:ElecStruc-E}
\end{figure}
\begin{figure}[tbp]
  \centering \includegraphics[width=15.0cm]{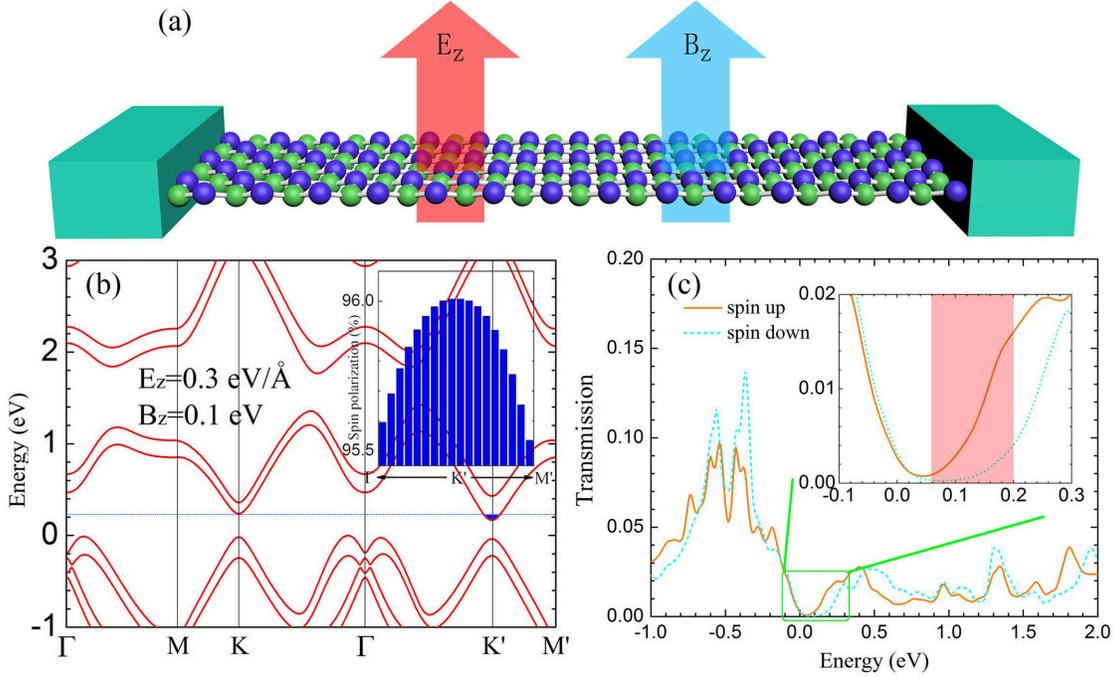}
  \renewcommand{\figurename}{Fig. S}
  \caption{(Color online) (a) Schematic model of a zigzag plumbene nanoribbon with two electrodes, perpendicular electric and magnetic fields. (b) The bulk band structure and (c) the spin-resolved transmission spectrum of the zigzag nanoribbon with $E_z=0.3$ eV/\AA \ and $B_z=0.1$ eV. The number of zigzag chains is $N_y = 6$ and the number of Pb atoms on each zigzag chain is $N_x = 20$ in (c). Inset of (b): the spin polarization corresponding to the blue area on the conduction band. Inset of (c): a partial enlargement in the energy range of $-0.1 \sim 0.3$ eV.}
  \label{fig:zz-nanoribbon}
\end{figure}

We have employed a sixteen-band tight-binding model to fit the band structure according to the first-principles result in the main text.
An electric field $E_z$ is assumed to be homogeneous along the $z$ direction.
The calculated band structure with $E_z=0.5$ eV/{\AA} is addressed in Fig. S\ref{fig:ElecStruc-E} (a). The perpendicular electric field results in a band splitting in the entire Brillouin zone except the time-reversal points $\Gamma$ and $M$.
With increasing $E_z$, the band splitting is increased. Meanwhile, the energy gap is decreased linearly, as plotted in Fig. S\ref{fig:ElecStruc-E} (d).
To explore and understand the properties of plumbene under perpendicular electric field, we next do a detailed analysis near the Fermi level around $K$ and $K'$ points. Fig. S\ref{fig:ElecStruc-E} (b) and (c) show spin-resolved band structures of plumbene and spin polarizations at $K$ and $K'$ points. The polarization of conduction band is much larger than that of valence band and up to $96 \%$ for $E_{z}=0.5$ eV/{\AA}.
Fig. S\ref{fig:ElecStruc-E} (e) shows the energy splittings of valence and conduction bands at $K$ point as functions of $E_{z}$. Both of them increase linearly with increasing the electric field strength. This directly leads to a linear reduction of the global gap (Fig. S\ref{fig:ElecStruc-E} (d)).
Fig. S\ref{fig:ElecStruc-E} (f) shows the evolution of the polarization of the conduction band at $K$ point with the perpendicular electric field $E_z$. The value of the polarization is almost decreased linearly, but it is still more than $95 \%$ when the field is up to $1.5$ eV/{\AA}.
We further consider an external magnetic field $B_z$ along the $z$ direction.
Fig. S\ref{fig:zz-nanoribbon} (b) shows the band structure with $E_{z}=0.3$ eV/{\AA} \ and $B_{z}=0.1$ eV. The corresponding spin polarization reaches $96 \%$ at the bottom of the lower-energy valley.

To theoretically investigate transport properties of the four-state spin-valley filter mentioned in the main text, we consider such a system which contains a plumbene nanoribbon with zigzag boundaries, left and right semi-infinite electrodes, as shown in Fig. S\ref{fig:zz-nanoribbon} (a). 
The retarded Green¡¯s function is defined as follows,
\begin{equation}
{G^R}\left( \varepsilon  \right){\rm{ = }}\frac{1}{{\left( {\varepsilon {\rm{ + }}i{\eta_{0} ^ + }} \right)I - {H_{Pb}} - {\Sigma _L} - {\Sigma _R}}},
\end{equation}
where ${H_{Pb}}$ is the Hamiltonian of the plumbene nanoribbon with external perpendicular fields $E_z$ and $B_z$, and $I$ is the identity matrix. ${\Sigma _{L/R}}$ are the self-energy operators and describe the effect of two semi-infinite electrodes. During the calculations, the wide-band approximation has been used and the real part of the self-energy is ignored\cite{JMMM350.6}. The imaginary part is set to $0.05$ eV. The Fisher-Lee relation $T\left( \varepsilon  \right) = Trace\left[ {{\Gamma _L}{G^R}\left( \varepsilon  \right){\Gamma _R}{G^A}\left( \varepsilon  \right)} \right]$ is used to calculate the transmission spectrum through the plumbene nanoribbon, where ${G^A}\left( \varepsilon  \right)$ is advanced Green¡¯s function and ${\Gamma _{L/R}} =  - 2{\mathop{\rm Im}\nolimits} {\Sigma _{L/R}}$\cite{PRB23.6851}.

Fig. S\ref{fig:zz-nanoribbon} (c) presents a spin-resolved transmission spectrum of the filter with  $E_z=0.3$ eV/\AA \ and $B_z=0.1$ eV. Actually, their values are very large. Here we choose them theoretically in order to display the external-field effects clearly and explicitly. The transmission coefficients are splitted due to the presence of external fields. In the range of $E = 0.06 \sim 0.2$ eV, the down-spin one is suppressed significantly and the up-spin one is relatively high. When the chemical potential is tuned into the energy range, it can achieve a spin-polarized transport. Moreover, according to the characteristic of the conduction band in Fig. S\ref{fig:zz-nanoribbon} (b), the valley polarization is also realized in this energy range. Therefore, a four-state spin-valley filter can be obtained through changing the directions of the applied fields, as described in the main text.

\begin{figure}[tbp]
  \centering \includegraphics[width=10.0cm]{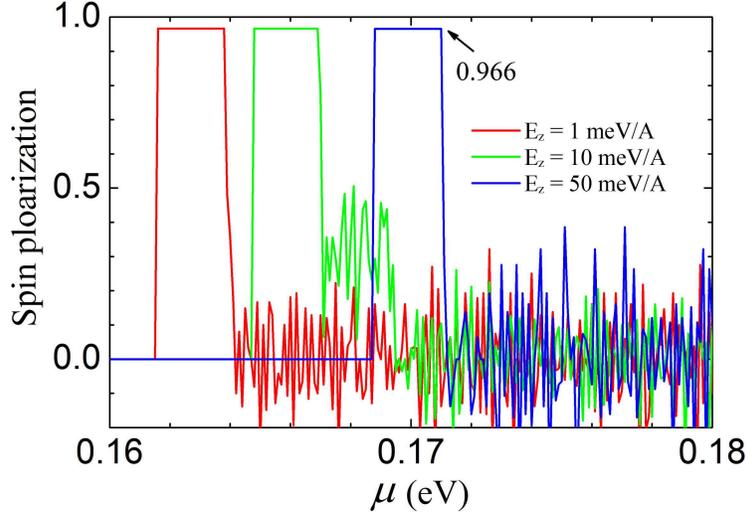}
  \renewcommand{\figurename}{Fig. S}
  \caption{(Color online) Spin polarizations as functions of chemical potential $\mu$ with $E_z = 1, 10, 50$ meV/\AA \ and $B_z = 1.16$ meV.}
  \label{fig:polar-Ez}
\end{figure}
As a supplement to Fig. 6 (b) of the main text, the spin polarizations with different electric field strengths are investigated in Fig. S\ref{fig:polar-Ez}. These field strengths can be realized in experiments. We find that both the spin polarization and the energy range of the polarization almost remain constant.